\documentclass[epj,nopacs]{svjour}
\usepackage{amssymb} 
\usepackage{graphicx} 
\begin{document}
\title{
\vspace*{-1.5\baselineskip}
\rightline{\textrm{\normalsize\rm CERN-TH/2003-253}}
The $\eta^\prime g^* g^{(*)}$ Vertex Function in 
Perturbative QCD \\ and $\eta^\prime$-Meson Mass Effects 
} 
\author{Ahmed Ali\inst{1}\thanks{
   On leave of absence from 
        Deutsches Elektronen-Syn\-chro\-tron 
        DESY, Hamburg, FRG.}%  
   \and Alexander Ya. Parkhomenko\inst{2}\thanks{
   On leave of absence from
        Department of Theoretical Physics,
        Yaroslavl State University,
        Sovietskaya~14, 150000 Yaroslavl, Russia.}%
}                     % Do not remove
\institute{ 
  Theory Division, CERN, CH-1211 Geneva 23, Switzerland 
  \and 
  Institut f$\ddot {\rm u}$r Theoretische Physik,
  Universit$\ddot {\rm a}$t Bern, CH-3012 Bern, Switzerland
}
\date{\phantom{Received: date / Revised version: date}} 
\abstract{
The $\eta^\prime g^* g^{(*)}$ effective vertex function (EVF) is
calculated in the QCD hard-scattering approach, taking into account
the $\eta^\prime$-meson mass. We work in the approximation in which
only one non-leading Gegenbauer moment in both the quark-antiquark 
and gluonic light-cone distribution amplitude for the 
$\eta^\prime$-meson is kept. The EVF with one off-shell gluon is shown 
to have the form $F_{\eta^\prime g^* g} (q_1^2, 0, m_{\eta^\prime}^2) 
= m_{\eta^\prime}^2 H(q_1^2)/(q_1^2 - m_{\eta^\prime}^2)$, valid 
for $|q_1^2| > m_{\eta^\prime}^2$. An interpolating formulae for 
the EVF in the space-like region of the virtuality~$q_1^2$, which 
satisfies the QCD-anomaly normalization for on-shell gluons and 
the perturbative-QCD result for the gluon virtuality 
$|q_1^2| \gtrsim 2$~GeV$^2$, is also presented. 
\PACS{
      {PACS-key}{discribing text of that key}   \and
      {PACS-key}{discribing text of that key}
     } % end of PACS codes
} %end of abstract
\authorrunning{A.~Ali, A.Ya.~Parkhomenko}
\titlerunning{
The $\eta^\prime g^* g^{(*)}$ vertex function in
perturbative QCD
}
\maketitle
\section{Introduction}
\label{sec:intro}

The $\eta^\prime g^* g^{(*)}$ effective vertex function (EVF) 
[or the $\eta^\prime$~-- gluon transition form factor], 
$F_{\eta^\prime g^* g^*} (q_1^2, q_2^2, m_{\eta^\prime}^2)$, 
enters in a number of decays such as $J/\psi \to \eta^\prime \gamma$, 
$\Upsilon \to \eta^\prime X$, $\Upsilon \to \eta^\prime \gamma$, 
$B \to (\pi, \rho, K, K^*) \, \eta^\prime$, $B \to \eta^\prime X_s$, 
and hadronic production processes, such as 
$N + N(\bar{N}) \to \eta^\prime X$, and hence is of great 
phenomenological importance.

At low gluon virtualities, $|q_1^2| \sim |q_2^2| \ll m_{\eta^\prime}^2$, 
the EVF is determined by the QCD anomaly while in inclusive decays 
of $B$- or $\Upsilon$-mesons, the gluon virtualities can be large enough 
($|q_i^2| \gtrsim m_{\eta^\prime}^2$) to allow the perturbative-QCD  
consideration for the EVF. Indeed, the hard part of the 
$\eta^\prime$-meson energy spectrum in the inclusive decay 
$\Upsilon (1S) \to g g g^* (g^* \to \eta^\prime g) \to \eta^\prime X$ 
is well reproduced when the per\-tur\-ba\-ti\-ve-QCD form for the 
$\eta^\prime$~-- gluon transition form factor is used in the 
analysis~\cite{Artuso:2002px,Kagan:2002dq,Ali:2003vw}.  

In the perturbative-QCD framework, the $\eta^\prime g^* g^{(*)}$ EVF 
can be calculated as a convolution of a hard-scattering kernel with 
the $\eta^\prime$-meson wave-function. For the energetic 
$\eta^\prime$-meson, transverse degrees of freedom of the 
meson constituents can be neglected and its wave-function is well 
described in terms of the quark-antiquark and gluonic light-cone 
distribution amplitudes (LCDAs). As the gluonic content of the 
$\eta^\prime$-meson is very important in many process, its effect 
can not be ignored. In this approach, the $\eta^\prime g^* g^{(*)}$ 
EVF has been studied by several 
groups~\cite{Muta:1999tc,Ali:2000ci,Kroll:2002nt,Agaev:2002ek}. 
As the $\eta^\prime$-meson mass is relatively large, it is not a 
good approximation to neglect it in certain kinematical regions, 
in particular, when the gluon virtuality is time-like. 
A consistent treatment of the $\eta^\prime$-meson mass effect 
in the EVF was undertaken by us recently~\cite{Ali:2003kg} and 
the results obtained are presented in this report.   

%
%%%%%%%%%%%%%%%%%%%%%%%%%%%%%%%%%%%%%%%%%%%%%%%%%%%%%%%%%%%%%%%%%%%%%%%%%
%
%\hspace{90mm}
\begin{figure*}[htb]%\sidecaption
\resizebox{0.60\textwidth}{!}{%
\includegraphics*{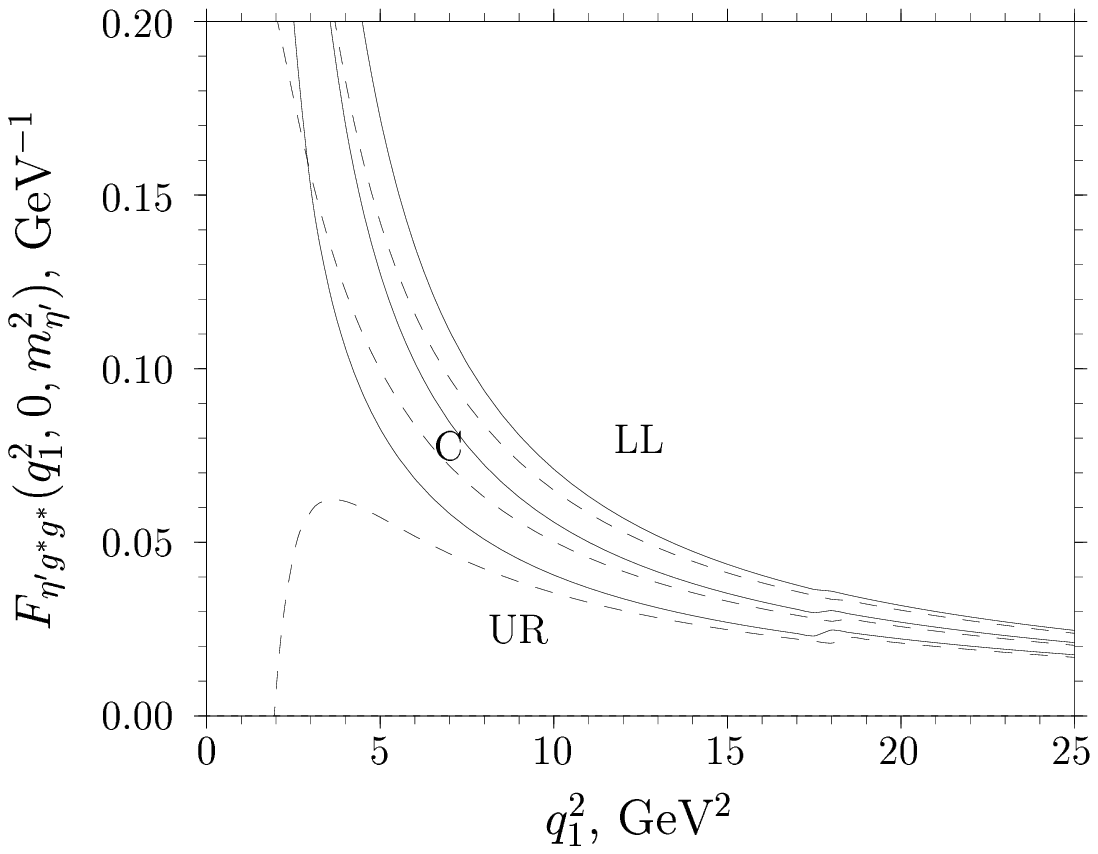}
\includegraphics*{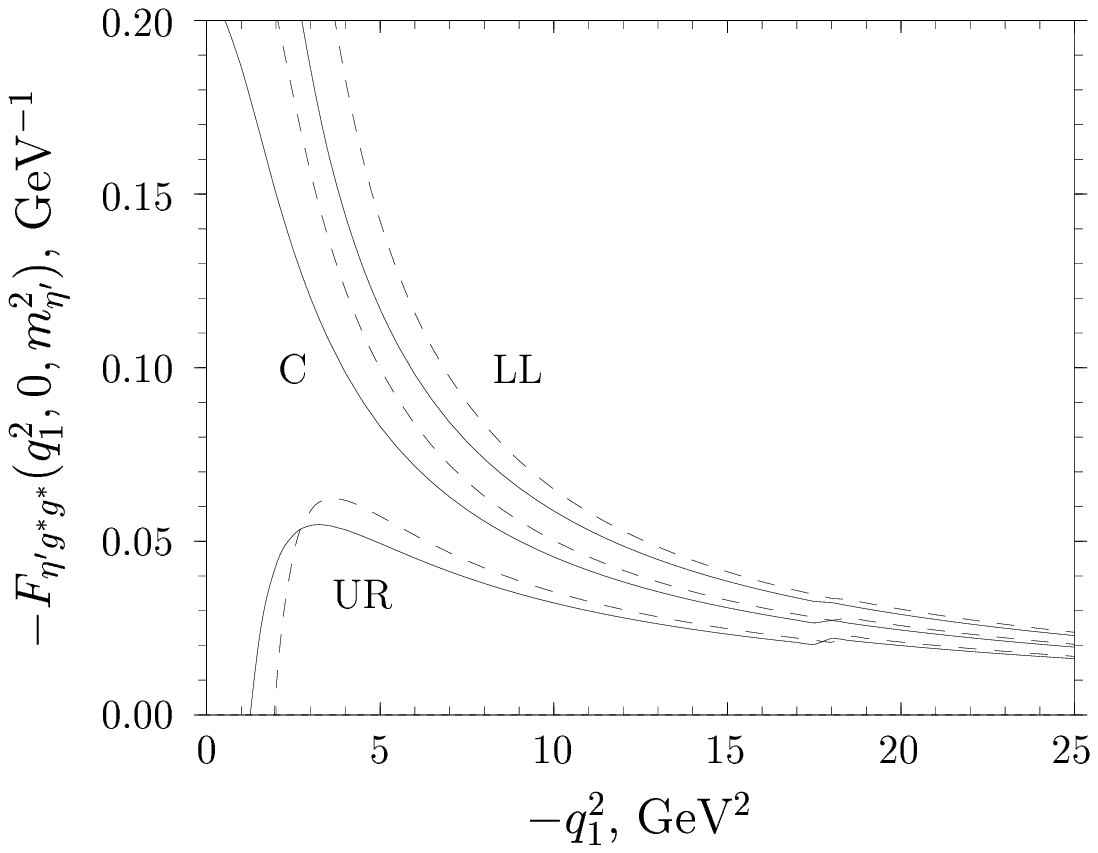}
}
\caption{
         The $\eta^\prime g^* g$ EVF for the time-like (left frame) 
         and space-like (right frame) gluon virtuality $q_1^2$ when 
         the second gluon is on the mass shell ($q_2^2 = 0$). The solid
         and dashed curves are plotted for the EVF with and without
         taking into account the $\eta^\prime$-meson mass, respectively.
         The labels~C, LL, and~UR correspond to the central, lower-left
         and upper-right points of the combined best fit of the
         Gegenbauer coefficients presented in Fig.~5 
         of~\cite{Ali:2003vw} and given in~(\ref{eq:combined-fit})
}
\label{fig:ff}
\end{figure*}
%
%%%%%%%%%%%%%%%%%%%%%%%%%%%%%%%%%%%%%%%%%%%%%%%%%%%%%%%%%%%%%%%%%%%%%%%%%
%

\section{Light-Cone $\eta^\prime$-Meson Wave-Function} 
\label{sec:wave-function}

The $\eta^\prime$-meson contains both the quark-antiquark 
and gluonic components and, for energetic $\eta^\prime$-meson, 
its wave-function can be presented in 
the form of the twist decomposition and described by the LCDAs. 
Here, we restrict ourselves to the leading-twist (twist-two) 
approximation only. An underlying theoretical basis for such 
a description can be found in~\cite{Kroll:2002nt,Ali:2003kg}.
The twist-two LCDAs are usually used in the following approximate forms: 
\begin{eqnarray} 
\phi^{(q)}_{\eta^\prime} (u, Q^2) & = & 6 u \bar u  
\left [ 1 + 6 (1 - 5 u \bar u) \, A_2 (Q^2) \right ] ,  
%\qquad 
\nonumber \\ 
\phi^{(g)}_{\eta^\prime} (u, Q^2) & = &  
5 u^2 \bar u^2 \, (u - \bar u) \, B_2 (Q^2) ,   
\label{eq:DAqg}
\end{eqnarray}
where only the second Gegenbauer moments~$A_2 (Q^2)$ 
and $B_2 (Q^2)$ are kept. As the quark-antiquark and gluonic 
components are mixed under the scale evolution, the Gegenbauer 
moments introduced above are superpositions of the non-perturbative 
parameters~$B^{(q)}_2 (\mu_0^2)$ and~$B^{(g)}_2 (\mu_0^2)$ 
[the Gegenbauer coefficients]: 
\begin{eqnarray} 
A_2 (Q^2) & = & B^{(q)}_2 \left [ 
\frac{\alpha_s (\mu_0^2)}{\alpha_s (Q^2)} \right ]^{\gamma_+^2}
\!\! + \rho_2^{(g)} \, B^{(g)}_2 \left [ 
\frac{\alpha_s (\mu_0^2)}{\alpha_s (Q^2)} \right ]^{\gamma_-^2} \!\! ,   
\quad 
\label{eq:An} \\
B_2 (Q^2) & = & \rho_2^{(q)} \, B^{(q)}_2 \left [ 
\frac{\alpha_s (\mu_0^2)}{\alpha_s (Q^2)} \right ]^{\gamma_+^2}
\!\! + B^{(g)}_2 \left [ 
\frac{\alpha_s (\mu_0^2)}{\alpha_s (Q^2)} \right ]^{\gamma_-^2} \!\! .    
\quad 
\label{eq:Bn} 
\end{eqnarray} 
The Gegenbauer coefficients at the initial scale~$\mu_0$ of the LCDA 
evolution can be estimated by non-perturbative methods or extracted 
from the experimental data sensitive to the internal structure of 
the $\eta^\prime$-meson.   
In particular, a fit to the CLEO and L3 data on the $\eta^\prime -
\gamma$ transition form factor for $Q^2 > 2$~GeV$^2$ was recently 
undertaken in~\cite{Kroll:2002nt}. 
The other process which allows to get an independent information
on the Gegenbauer coefficients in the $\eta^\prime g^* g$ EVF is 
the inclusive $\Upsilon(1S) \to \eta^\prime X$ decay.
The $\eta^\prime$-meson energy spectrum in this decay was
recently measured by the CLEO collaboration~\cite{Artuso:2002px} 
and the data on the hard part of the spectrum are in agreement
with the perturbative-QCD analysis~\cite{Kagan:2002dq,Ali:2003vw}.
Current experiments and the theoretical analysis undertaken 
in these processes individually leave an order of magnitude
uncertainties on the Gegenbauer coefficients but the combined  
fit of the data allows to reduce substantially these uncertainties 
and results in the following values~\cite{Ali:2003vw}: 
\begin{eqnarray}
B_2^{(q)} (2~{\rm GeV}^2) & = & - 0.008 \pm 0.054,  
\label{eq:combined-fit} \\ 
B_2^{(g)} (2~{\rm GeV}^2) & = & 4.6 \pm 2.5. 
\nonumber 
\end{eqnarray}
%

%
%%%%%%%%%%%%%%%%%%%%%%%%%%%%%%%%%%%%%%%%%%%%%%%%%%%%%%%%%%%%%%%%%%%%%%%%%
%
%\vspace{-15mm}
\begin{figure*}[htb]%\sidecaption
\resizebox{0.6\hsize}{!}{%
\includegraphics*[height=0.25\textheight,width=0.25\textwidth]
      {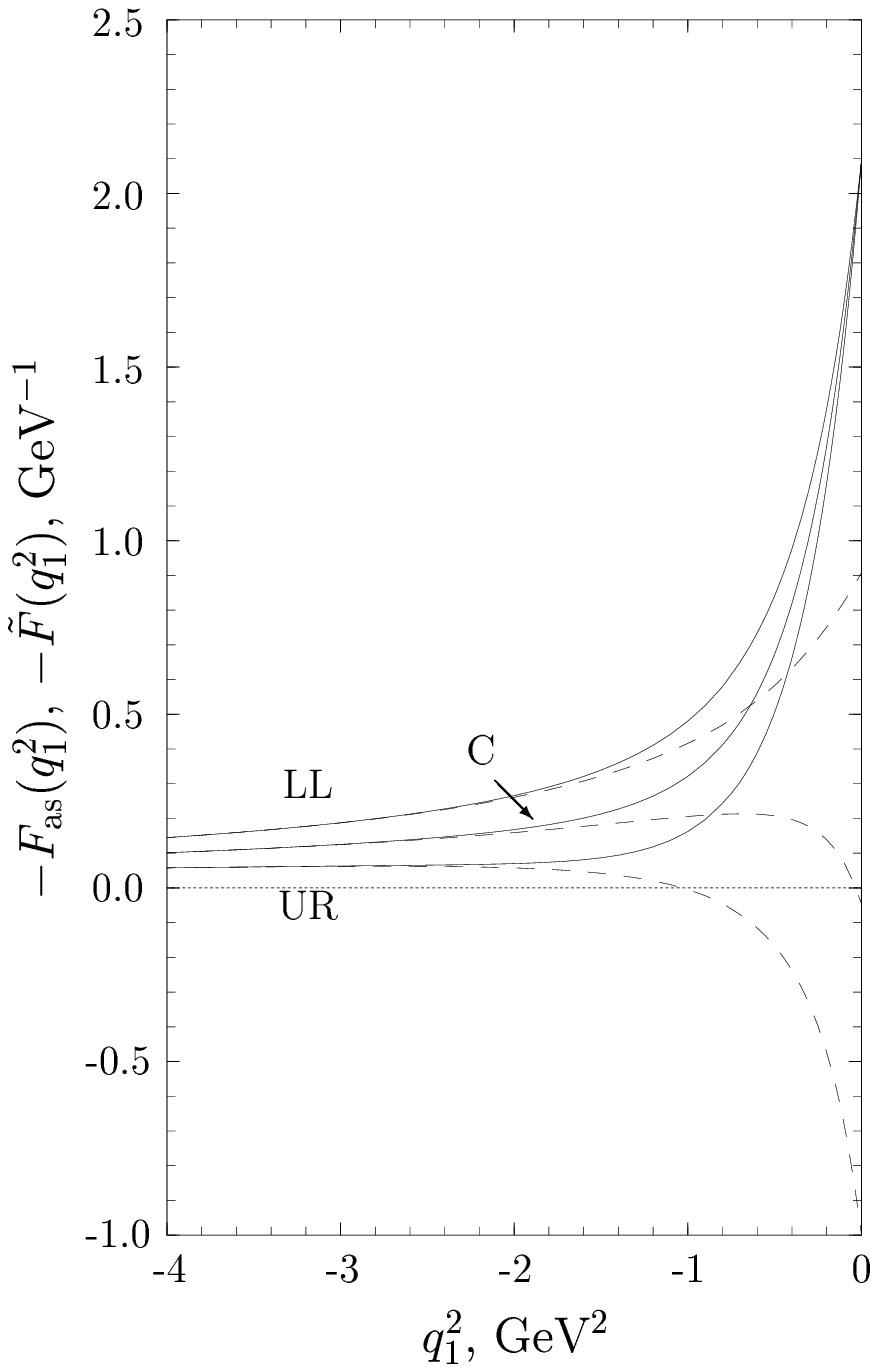}
\includegraphics*[height=0.25\textheight,width=0.25\textwidth]
      {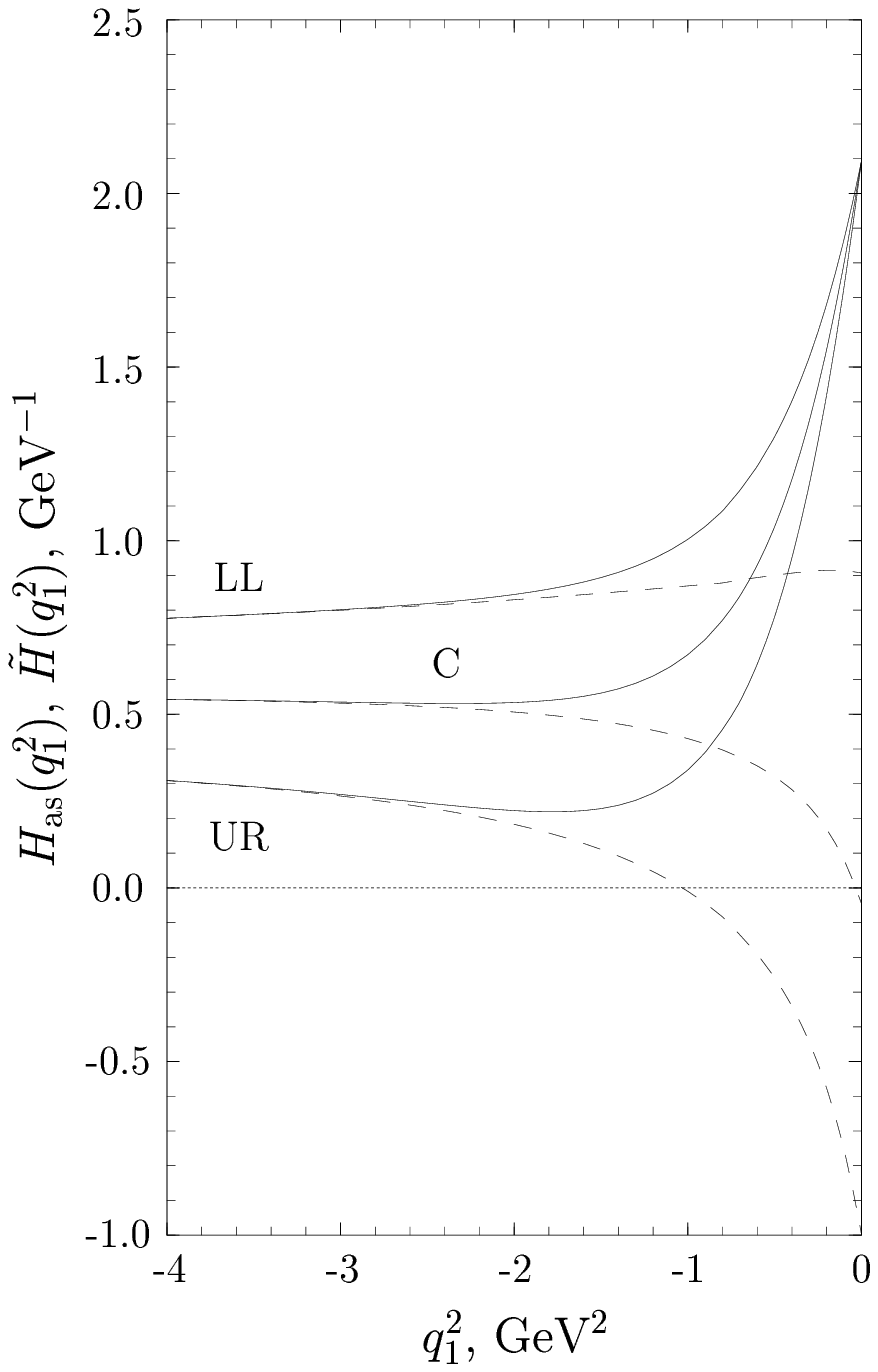}
}
\caption{
         The $\eta^\prime - g$ transition form factor in the
         perturbative-QCD approach and using the interpolating
         formulae for the space-like region of the gluon virtuality
         $q_1^2$. The left frame shows the functions
         $F_{\rm as}(q_1^2)$ (dashed curves) and $\tilde{F}(q_1^2)$
         (solid curves). The right frame shows the functions
         $H_{\rm as} (q_1^2)$ (dashed curves) and $\tilde H (q_1^2)$
         (solid curves) which are connected with $F_{\rm as}(q_1^2)$ 
         and $\tilde{F}(q_1^2)$ by~(\ref{eq:F-mass-shell}). 
%         The functions $F_{\rm as}(q_1^2)$ and 
%         $H_{\rm as} (q_1^2)$ [$\tilde{F}(q_1^2)$ and $\tilde H
%         (q_1^2)$] are connected by~(\ref{eq:F-mass-shell}).
         The labels on the curves are the same as in Fig.~\ref{fig:ff}
}
\label{fig:asym}
\end{figure*}
%
%%%%%%%%%%%%%%%%%%%%%%%%%%%%%%%%%%%%%%%%%%%%%%%%%%%%%%%%%%%%%%%%%%%%%%%%%
%

\section{The $\eta^\prime g^* g^{(*)}$ Effective Vertex Function} 
\label{sec:EVF}

In the momentum space, the $\eta^\prime g^* g^*$ EVF can be 
extracted from the invariant matrix element of the process 
$\eta^\prime (p) \to g^* (q_1) \, g^* (q_2)$: 
\begin{equation}
{\cal M} \equiv 
F_{\eta^\prime g^* g^*} (q_1^2, q_2^2, m_{\eta^\prime}^2) \,
\delta_{A B} \, \varepsilon^{\mu \nu \rho \sigma} \,
\varepsilon^{A*}_{1\mu} \varepsilon^{B*}_{2\nu} q_{1\rho} q_{2\sigma} . 
\label{eq:EV-def}
\end{equation}
This amplitude gets contributions from both the quark-antiquark and 
gluonic components of the $\eta^\prime$-meson and the corresponding 
individual amplitudes~${\cal M}^{(q)}$ and~${\cal M}^{(g)}$ can be 
calculated as follows: 
\begin{eqnarray}
{\cal M}^{(q)} & = & i f_{\eta^\prime} 
\int\limits_0^1 du \, \phi^{(q)}_{\eta^\prime} (u, Q^2) \, 
{\cal P}^{(q)}_{j \beta b; i \alpha a} \, 
\delta^{a b} \big [ T^{(q)}_{\rm H} \big ]^{\alpha \beta}_{i j} ,  
\quad 
\label{eq:MEqq-def} \\
{\cal M}^{(g)} & = &   
\frac{i f_{\eta^\prime}}{2} \int\limits_0^1 du \, 
\frac{\phi^{(g)}_{\eta^\prime} (u, Q^2)}{u \bar u} \,
{\cal P}^{(g)}_{\sigma D; \rho C} \,
\big [ T^{(g)}_{\rm H} \big ]^{\rho \sigma}_{C D} . 
\quad 
\label{eq:MEgg-def}  
\end{eqnarray}
Here, $\big [ T^{(q)}_{\rm H} \big ]^{\alpha \beta}_{i j}$ 
($\big [ T^{(g)}_{\rm H} \big ]^{\rho \sigma}_{C D}$)  
is the quark (gluonic) hard-scattering kernel calculated 
in the perturbative QCD and ${\cal P}^{(q)}_{j \beta b;i \alpha a}$  
(${\cal P}^{(g)}_{\mu A; \nu B}$) is the $\eta^\prime$-meson 
projection operator onto the quark-antiquark (two-gluon) 
state~\cite{Ali:2003vw}.  
 
The results for the quark-antiquark and gluonic parts of 
the $\eta^\prime g^* g^*$ EVF can be written in the form:  
\begin{eqnarray}
F^{(q)}_{\eta^\prime g^* g^*} (q^2, \omega, \eta) & = & 
\frac{4 \pi \alpha_s (Q^2)}{m_{\eta^\prime}^2 \, \lambda} \, 
\frac{3 f_{\eta^\prime} \sqrt{N_f}}{N_c}
\label{eq:epgg-res-gen} \\ 
& \times &  
\left \{ G^{(q)}_0 (\omega, \eta) + 6 A_2 (Q^2) G^{(q)}_2 (\omega, \eta)
\right \} ,
\nonumber 
\end{eqnarray}
\begin{equation}
F^{(g)}_{\eta^\prime g^* g^*} (q^2, \omega, \eta) =  
- \frac{4 \pi \alpha_s (Q^2)}{m_{\eta^\prime}^2 \lambda} \, 
\frac{5 f_{\eta^\prime}}{2 \sqrt{N_f}} \, B_2 (Q^2) \, 
{\rm G}^{(g)}_2 (\omega, \eta) ,  
\label{eq:GFF-result} 
\end{equation}
where the following kinematical quantities: $q^2 = q_1^2 + q_2^2$,
$\omega = (q_1^2 - q_2^2)/q^2$, $\eta = m_{\eta^\prime}^2/q^2$,
and the parameter $\lambda = \sqrt{1 - \eta (2 - \eta)/\omega^2}$
are introduced. The explicit forms of the functions 
$G^{(q)}_0 (\omega, \eta)$, ${\rm G}^{(q)}_2 (\omega, \eta)$, 
and~$G^{(g)}_2 (\omega, \eta)$ can be found in~\cite{Ali:2003kg}.  
When the $\eta^\prime$-meson mass is neglected ($m_{\eta^\prime} = 0$), 
the usual $1/q^2$ behavior of the EVF is reproduced~\cite{Ali:2003kg}.   
Both contributions contain the factor $1/\lambda$ which is equal to  
$m_{\eta^\prime}^2/(q_1^2 - m_{\eta^\prime}^2)$ for the case when the 
second gluon in the final state is on the mass shell ($q_2^2 = 0$). 
Thus, the phenomenological form for the 
$\eta^\prime - g$ transition form factor suggested by Kagan 
and Petrov~\cite{Kagan:1997qn}: 
\begin{equation} 
F (q_1^2) \equiv 
F_{\eta^\prime g^* g} (q_1^2, 0, m_{\eta^\prime}^2) = 
\frac{m_{\eta^\prime}^2 \, H(q_1^2)}{q_1^2 - m_{\eta^\prime}^2} , 
\label{eq:F-mass-shell}
\end{equation} 
is naturally reproduced in the LCDA approach when the
$\eta^\prime$-meson mass is taken into account. As a disadvantage 
of this approach, one encounters a singularity at $q_1^2 =
m_{\eta^\prime}^2$ which, however, can be removed by an inclusion 
of transverse degrees of freedom for the partons in the 
$\eta^\prime$-meson. In contrast to~\cite{Kagan:1997qn} 
where it was suggested to approximate the function $H(q_1^2)$ 
in~(\ref{eq:F-mass-shell}) by a constant value,  
$H_0 = 1.7$~GeV$^{-1}$, the explicit form of this function 
was calculated by us in the framework of the QCD hard-scattering 
approach~\cite{Ali:2003kg}. It can be presented the following 
approximate form: 
\begin{equation} 
H_{\rm as} (q_1^2) = 
\frac{4 \pi \alpha_s (Q^2)}{m_{\eta^\prime}^2} \, 
\sqrt 3 f_{\eta^\prime} \left [ 1 + A_2 (Q^2) -  
\frac{5}{36} \, B_2 (Q^2) \right ] ,    
\label{eq:H-mass-shell}
\end{equation} 
in the region of the applicability of the perturbative QCD which 
was taken as $q_1^2 < - 1$~GeV$^2$ and $q_1^2 > 2$~GeV$^2$. 
The dependence on~$q_1^2$ in the r.h.s. of~(\ref{eq:H-mass-shell}) 
is coming only through the combination 
$Q^2 = |q_1^2| + m_{\eta^\prime}^2$.   

The dependence of the EVF on the gluon virtuality~$q_1^2$ in the 
time- and space-like regions for the combined best-fit 
values~(\ref{eq:combined-fit}) of the Gegenbauer coefficients is 
presented in Fig.~\ref{fig:ff}. The inclusion of the 
$\eta^\prime$-meson mass reduces the parametric dependence on 
the Gegenbauer coefficients of the $\eta^\prime g^* g$ EVF 
in the time-like region of the gluon virtuality. This is generally 
not the case for the space-like gluon virtuality, in particular for 
the low values, as the $\eta^\prime$-meson mass effects are not 
as pronounced. 

A {\it formal} limit of the function~(\ref{eq:H-mass-shell}) for 
on-shell gluons ($q_1^2 = 0$) exists with a strong dependence on 
the Gegenbauer coefficients as shown in Fig.~\ref{fig:asym}. 
It is well known that the value of the $\eta^\prime g^* g^*$ EVF 
in this limit is determined by the anomaly: 
\begin{equation}  
F^{\rm A}_{\eta^\prime gg} = 
- 4 \pi \alpha_s (m_{\eta^\prime}^2) \,
\frac{1}{2 \pi^2 f_{\eta^\prime}} = - H_{\rm A} ,   
\label{eq:PCAC-value}
\end{equation}
which is substantially different from the limiting values of the 
perturbative-QCD motivated EVF (see Fig.~\ref{fig:asym}). 
Thus, the behavior of~$H_{\rm as} (q_1^2)$ should be modified 
accordingly. In~\cite{Ali:2003kg}, we suggested the following 
interpolating formulae:  
\begin{equation}
\tilde H (q_1^2) = H_{\rm as} (q_1^2) + 
\left [ H_A - H_{\rm as} (0) \right ] 
\exp \left [ C_s \, q_1^2/m_{\eta^\prime}^2 \right ] , 
\label{eq:H-02}
\end{equation}
to improve the EVF behavior at the low space-like virtuality. In the 
numerical analysis, $C_s = 2$ was adopted for the free parameter 
introduced, which allows a smooth interpolation between the anomaly 
normalization at $q_1^2 = 0$ and the perturbative-QCD result for 
the large-$|q_1^2|$ region. As seen from Fig.~\ref{fig:asym}, this   
interpolation strongly decrease the dispersion in the region of the 
low space-like gluon virtuality.

\section{Summary}  
\label{sec:summary} 

The $\eta^\prime g^* g^{(*)}$ EVF is calculated in the perturbative-QCD 
approach using the LCDAs for the $\eta^\prime$-meson wave-function 
with the inclusion of the $\eta^\prime$-meson mass. 
If one of the gluons is on the mass shell, the pole-like behavior
of the $\eta^\prime$~-- gluon transition form factor emerges in this 
approach for both the quark-antiquark and gluonic parts of the EVF.   
The corresponding function~$H (q_1^2)$ is perturbatively calculated. 
The $\eta^\prime$-meson mass effects are analyzed numerically with the 
result that they are important for lower values of the gluon virtuality, 
in particular in the time-like region. 
An interpolating formulae connecting the QCD-anomaly value and 
the perturbative-QCD behavior of the $\eta^\prime$~-- gluon 
transition form factor is presented for the space-like region 
of the gluon virtuality, taking into account the 
$\eta^\prime$-meson mass, which modifies the EVF significantly 
in the region $\vert q_1^2 \vert < 1$ GeV$^2$ and reduces also  
the theoretical dispersion in this region considerably. 

\begin{acknowledgement}
The work of A.Ya.P. has been
supported by the Schweizerischer Nationalfonds.
\end{acknowledgement}

\end{document}